\documentclass[reprint, prb, superscriptaddress, floatfix, noeprint]{revtex4-2}

\usepackage{stix}

\usepackage[dvipsnames]{xcolor}
\usepackage{hyperref}
\hypersetup{
    colorlinks=true,
    linkcolor=BlueViolet,
    citecolor=Blue,
    urlcolor=BlueViolet,
    pdfpagemode=FullScreen,
    }
\usepackage{graphicx}
\usepackage{placeins}
\usepackage{amsmath}
\usepackage{mathtools}
\usepackage{verbatim}
\usepackage{bm}
\usepackage{textcomp}
\usepackage[export]{adjustbox}
\usepackage[capitalize]{cleveref}
\usepackage{float}
\usepackage{braket}
\usepackage{tcolorbox}
\tcbuselibrary{theorems}
\usepackage{tikz}

\newcommand{\xRPA}{\textit{x}RPA}
\newcommand{\dRPA}{\textit{d}RPA}
\newcommand{\xdRPA}{\textit{xd}RPA}
\newcommand{\bvec}[1]{\boldsymbol{#1}}
\DeclareMathOperator{\Tr}{\mathrm{Tr}}

\usepackage{makerobust}
\newcommand{\makeauthor}[2]{\newcommand{#1}[1]{{%
  \protect%
  \color{#2}{%
    \bfseries\begingroup\escapechar=-1\edef\x{\endgroup\string#1}\x:%
  } ##1}}%
  \MakeRobustCommand#1}
\makeauthor{\lk}{orange}
\makeauthor{\af}{red}
\makeauthor{\ar}{green}

\makeatletter
\def\maketitle{
\@author@finish
\title@column\titleblock@produce
\suppressfloats[t]}
\makeatother
\newcommand{\supplement}[1]{%
\clearpage
\title{#1}
\maketitle
\setcounter{equation}{0}
\setcounter{figure}{0}
\setcounter{table}{0}
\setcounter{page}{1}
\makeatletter
\renewcommand{\thesection}{S\arabic{section}}
\renewcommand{\thesubsection}{\Alph{subsection}}
\renewcommand{\theequation}{S\arabic{equation}}
\renewcommand{\thefigure}{S\arabic{figure}}
\renewcommand{\thetable}{S\Roman{table}}
\renewcommand{\thepage}{S\arabic{page}}
\makeatother
\onecolumngrid}

\begin{document}
\title{Spin and Charge Fluctuation Induced Pairing in ABCB Tetralayer Graphene}

\author{Ammon Fischer}
\thanks{These authors contributed equally}
\affiliation{Institute for Theory of Statistical Physics, RWTH Aachen University, and JARA Fundamentals of Future Information Technology, 52062 Aachen, Germany}
\author{Lennart Klebl} 
\thanks{These authors contributed equally}
\affiliation{I. Institute for Theoretical Physics, Universität Hamburg, Notkestraße 9-11, 22607 Hamburg, Germany}
\author{Jonas B.~Hauck} 
\affiliation{Institute for Theory of Statistical Physics, RWTH Aachen University, and JARA Fundamentals of Future Information Technology, 52062 Aachen, Germany}
\author{Alexander Rothstein} 
\affiliation{2nd Institute of Physics and JARA-FIT, RWTH Aachen University, 52074 Aachen, Germany}
\affiliation{Peter Gr\"unberg Institute (PGI-9), Forschungszentrum J\"ulich, 52425 J\"ulich, Germany}
\author{Lutz Waldecker} 
\affiliation{2nd Institute of Physics and JARA-FIT, RWTH Aachen University, 52074 Aachen, Germany}
\author{Bernd Beschoten}
\affiliation{2nd Institute of Physics and JARA-FIT, RWTH Aachen University, 52074 Aachen, Germany}
\author{Tim O.~Wehling}
\affiliation{I. Institute for Theoretical Physics, Universität Hamburg, Notkestraße 9-11, 22607 Hamburg, Germany}
\affiliation{The Hamburg Centre for Ultrafast Imaging, 22761 Hamburg, Germany}
\author{Dante M.~Kennes}
\email{dante.kennes@mpsd.mpg.de}
\affiliation{Institute for Theory of Statistical Physics, RWTH Aachen University, and JARA Fundamentals of Future Information Technology, 52062 Aachen, Germany}
\affiliation{Max Planck Institute for the Structure and Dynamics of Matter, Center for Free Electron Laser Science, 22761 Hamburg, Germany}

\date{\today}
\begin{abstract}
Motivated by the recent experimental realization of ABCB stacked tetralayer graphene [Wirth \emph{et al.}, ACS Nano 16, 16617 (2022)], we study correlated phenomena in moir\'e-less graphene tetralayers for realistic interaction profiles using an orbital resolved random phase approximation approach. We demonstrate that magnetic fluctuations originating from local interactions are crucial close to the van Hove singularities on the electron- and hole-doped side promoting layer selective ferrimagnetic states. Spin fluctuations around these magnetic states enhance unconventional spin-triplet, valley-singlet superconductivity with $f$-wave symmetry due to intervalley scattering. Charge fluctuations arising from long range Coulomb interactions promote doubly degenerate $p$-wave superconductivity close to the van Hove singularities. At the conduction band edge of ABCB graphene, we find that both spin and charge fluctuations drive $f$-wave superconductivity. Our analysis suggests a strong competition between superconducting states emerging from  long- and short-ranged Coulomb interactions and thus stresses the importance of microscopically derived interaction profiles to make reliable predictions for the origin of superconductivity in graphene based heterostructures.
\end{abstract}

\maketitle

\paragraph*{Introduction.}
The experimental discovery of cascades of correlated phases and superconductivity in 
ultra-clean graphene multilayers without~\cite{seiler2022quantum,zhou2021superconductivity,holleis2023ising, zhou2022isospin,zhang2023enhanced, winterer2023ferroelectric} and with~\cite{cao_unconventional_2018, yankowitz2019tuning, lu2019superconductors, saito2020independent, stepanov2020untying, oh2021evidence, cao2021nematicity,
zhang2021ascendance, park2021tunable, cao2021pauli, kim2022evidence, liu2022isospin, kennes_moire_2021} a stacking twist has led to tremendous research interest. Experimental studies of multilayer graphene suggest the existence of ferromagnetic phases in the form of half- and quarter metals~\cite{zhou2021half,seiler2022quantum}, Wigner crystals~\cite{seiler2022quantum} and in particular superconductivity that emerges in Bernal bilayer (AB) and rhombohedral trilayer (ABC) graphene for external displacement fields and either in the presence of external magnetic fields~\cite{zhou2021superconductivity,zhou2022isospin} or proximity induced spin-orbit coupling~\cite{zhang2023enhanced,holleis2023ising}. 
From a fabrication point of view, untwisted graphene stacks are simpler to handle as twist angle variations and other stacking imperfections are easier to control. It is therefore that experiments along with atomistic theoretical studies can provide a microscopic understanding of correlation effects, in particular as hindering disorder effects can be disregarded and multilayer graphene stacks omit nm-scale unit cells as well as a more sophisticated theory of topological obstruction of the low-energy bands when compared to their twisted counterparts~\cite{ahn2019failure,song2019all,zou2018band}. 

Nonetheless band flattening in untwisted graphene stacks is restricted to small Fermi surface patches around the valleys $K$ and $K'$, which hampers theoretical descriptions without losing contact to the microscopics.
Previous works~\cite{chatterjee2022inter,szabo2022competing,szabo2022metals,PhysRevB.105.134524,li2023charge,pantaleon2022superconductivity,pantaleon2023superconductivity,cea2022superconductivity,jimeno2023superconductivity,PhysRevB.107.L041111,dai2022quantum,ghazaryan2023multilayer,wei2023,wagner2023superconductivity} resort to effective continuum or adapted Slonzecewski-Weiss-McClure models with ultra-violet cutoff that are only valid near the valleys $K^{(\prime)}$ and hence severely complicate a description of realistic orbital-resolved interaction profiles that account for short and long-ranged interactions consistently. Instead, an approximate spin/valley $SU(4)$ symmetry arises from considering the long-ranged tail of the Coulomb interaction alone. 

Motivated by the recent experimental characterization of different tetralayer graphene stacks by Wirth \emph{et al.}~\cite{Wirth2022Oct}, we remedy the aforementioned shortcomings and study correlated phenomena in moir\'e-less graphene tetralayers for \emph{ab-initio} motivated models and realistic interaction profiles using an orbital resolved random phase approximation approach (RPA). In particular, we focus on ABCB stacked graphene that breaks inversion symmetry and thus features intrinsic electrical fields. The latter open a band gap at charge neutrality and hence play the same role as external displacement fields that are necessary to stabilize superconductivity in bi- and trilayer graphene. While previous works mainly focused on pairing mediated by electron-phonon coupling~\cite{PhysRevB.105.L100503,PhysRevB.106.L180502,PhysRevB.106.024507}, fluctuations around flavor symmetry broken phases~\cite{chatterjee2022inter,szabo2022competing,szabo2022metals,PhysRevB.105.134524} or charge fluctuations arising from repulsive Coulomb interactions~\cite{li2023charge,pantaleon2022superconductivity,pantaleon2023superconductivity,cea2022superconductivity,jimeno2023superconductivity,PhysRevB.107.L041111,dai2022quantum,ghazaryan2023multilayer,wei2023,wagner2023superconductivity}, we discuss the interplay of spin and charge fluctuations to the formation of weak coupling instabilities on general grounds based on a miscroscopic model of the carbon $p_z$-orbitals. In view of the various spin-polarized phases governing large areas of the phase diagram of bi- and trilayer graphene~\cite{seiler2022quantum}, we first characterize the role of spin-fluctuations towards the formation of magnetic order in ABCB graphene. To disentangle the influence of long- and short-ranged interactions as well as their contributions to spin- and charge enhanced superconductivity, we next study three different pairing mechanisms based on (i) spin-fluctuations from local interactions ($x$RPA) (ii) screening of long-ranged Coulomb interactions ($d$PRA) and (iii) a combined approach ($xd$RPA) that captures both long and short ranged Coulomb interactions.

\paragraph*{Microscopic Model.}
We model the electronic band structure of graphene tetralayers employing a modified Slonzecewski-Weiss-McClure Hamiltonian~\cite{Rubio_flg_Tb}. The resulting kinetic Hamiltonian represents an atomistic model of tetralayer graphene consistent with first principle calculations~\cite{aoki_dependence_2007}, see Supplementary Material~(SM)~\cite{SM} for details on the parametrization.
\begin{figure}
    \centering
    \includegraphics[width=\columnwidth]{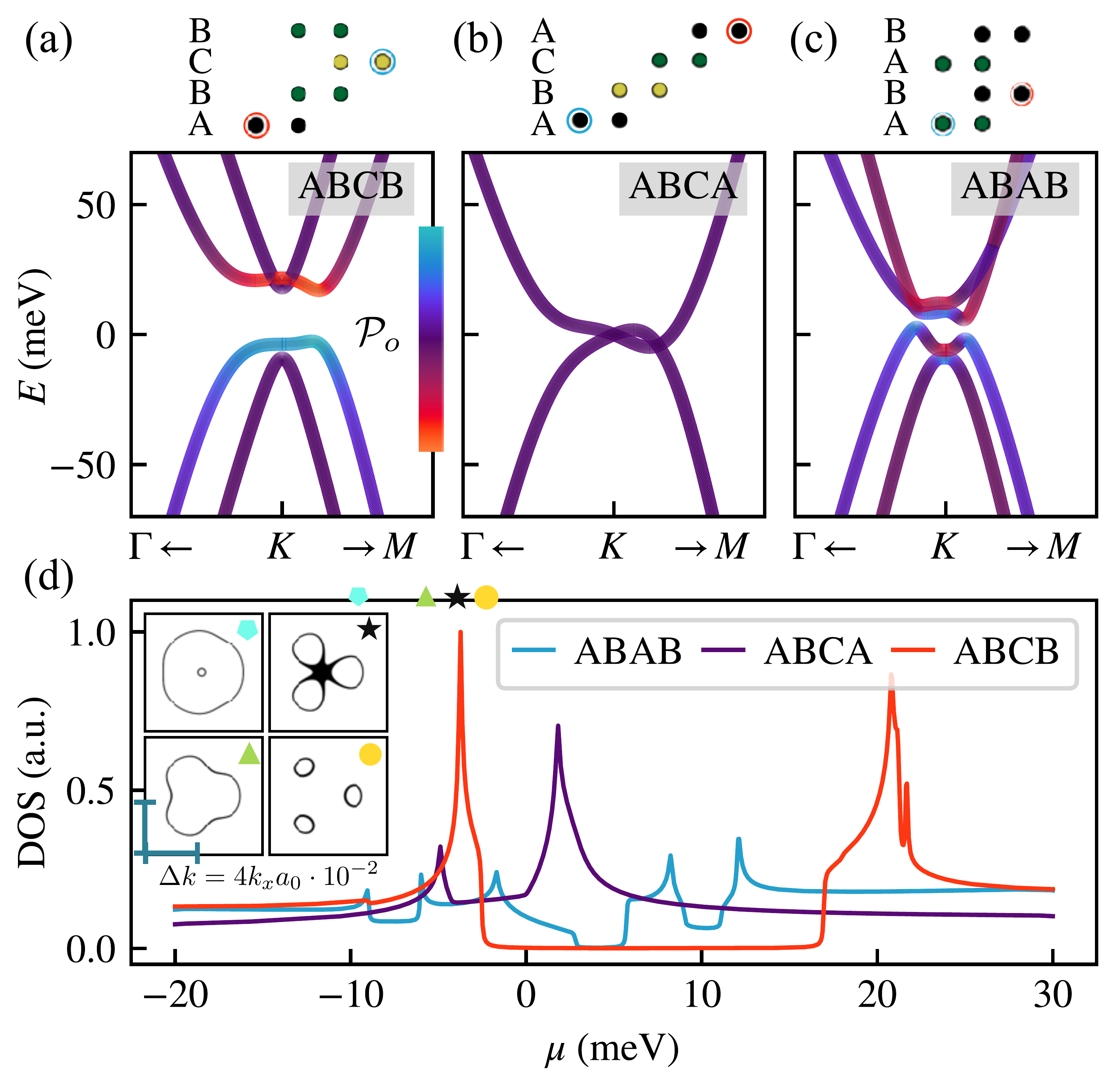}
    \caption{Band structure and density of states (DOS) for different tetralayer graphene structures without an applied electric field. Panels~(a-c): Low-energy band structure near valley $K$ with the relative orbital polarization $\mathcal{P}_o$ of the bands indicated by the diverging colormap. The different stacking configurations of ABCB, ABCA and ABAB tetralayer graphene are shown above while the two sites with dominant spectral weight are indicated. Panel~(d): DOS within the range of the valley-flat bands. Intrinsic electrical fields caused by broken inversion symmetry in ABCB stacked graphene open a bandgap of $\Delta_g \approx 20 \, \text{meV}$ and flatten the low-energy bands around the valleys $K,K'$. Hence, ABCB has the highest DOS near the VHS on the electron- and hole-doped side. The inset highlights different Fermi surfaces around valley $K$ as the chemical potential is tuned through the VHS on the hole-doped side. Starting from the edge of the valence band, the Fermi surface undergoes a Lifshitz transition as the the three pockets originating from trigonal warping continuously transform to an annular Fermi surface.}
    \label{fig:bands}
\end{figure}
The upper panels of \cref{fig:bands} give an overview of the non-interacting electronic structure of the three inequivalent thermally stable stacking configurations of tetralayer graphene: ABCB~(a), ABCA~(b), and ABAB~(c). We plot the low-energy band structure along the two irreducible directions in vicinity of the valley $K$. The color encodes the polarization $\mathcal{P}_o$ of the dominant orbitals. For ABCB graphene we find that most weight is concentrated on the a-site in the first layer and the b-site in the third layer. Due to broken layer inversion symmetry, ABCB graphene features a band gap of $\Delta_g \approx 20$ meV at charge neutrality without the application of an external electric field.  This is reflected in the density of states [DOS; see \cref{fig:bands}~(d)] that indicates van Hove singularities (VHS) close to charge neutrality on the electron- and hole-doped side. As the internal electric field in ABCB causes valley band flattening, the VHS is stronger in ABCB compared to ABCA and ABAB graphene. Therefore we expect the effects of electron-electron interactions to be most significant in ABCB graphene.  By inspection of the ABCB Fermi surface for various fillings we observe a Lifshitz transition from a (annular) single-pocket to a three-pocket structure, which signals an intriguing interplay between nesting- and DOS-driven Fermi surface instabilities and resembles AB graphene under an electrical field. As the latter hosts magnetic states~\cite{seiler2022quantum}, we start with a discussion of spin fluctuation induced instabilities in the following.

\begin{figure}
    \centering
    \includegraphics{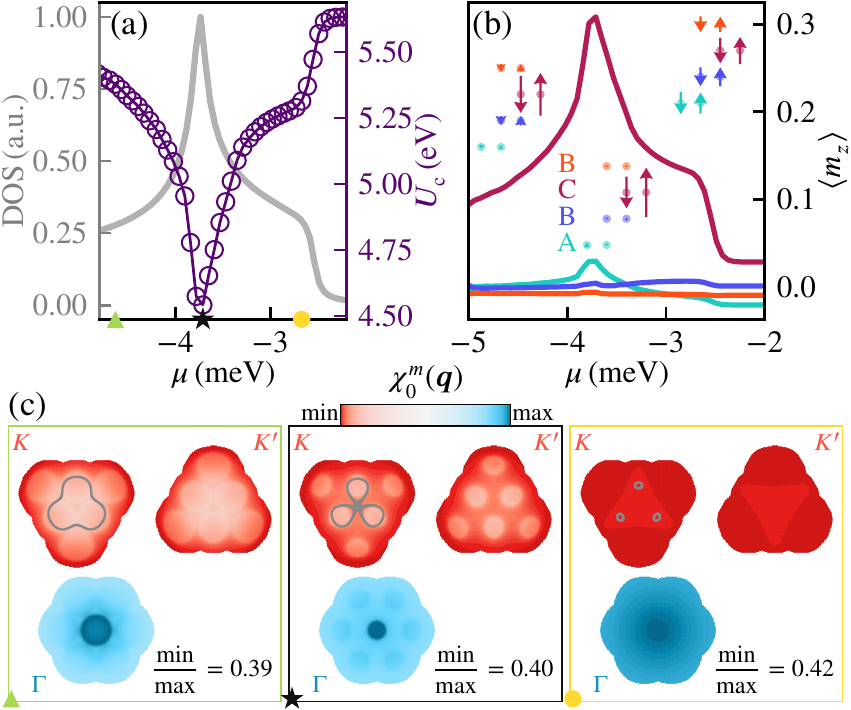}
    \caption{Spin correlations in the presence of local interactions in ABCB tetralayer graphene. Panel~(a) shows the critical on-site interaction $U_\mathrm{c}$ required for the onset of magnetic order for $\mu$ close to the van-Hove filling of the valence band (purple). We additionally include the density of states (DOS, gray). We expect an increased tendency towards magnetic order at the van-Hove filling $\mu_\mathrm{VHS}\approx-3.9\,\mathrm{meV}$. The leading magnetic correlations are displayed in panel~(b): Each curve corresponds to the magnetization within one layer. The insets depict the three different magnetic states found: At $\mu\approx\mu_\mathrm{VHS}$, local interactions drive ferrimagnetism in the C-layer (3). At $\mu<\mu_\mathrm{VHS}$, antiferromagnetic fluctuations in layer 3 are leading, and for states close to charge neutrality ($\mu>\mu_\mathrm{VHS}$) the antiferromagnetic fluctuations extend over all layers. In panel~(c) we show the momentum structure of the leading eigenvalue of the susceptibility matrix $\hat\chi_0(\bvec q)$ for $\mu<\mu_\mathrm{VHS}$, $\mu\approx\mu_\mathrm{VHS}$, and $\mu>\mu_\mathrm{VHS}$ [c.f. markers in~(a)]. The transfer momenta supported by the Fermi surface are $\bvec q\approx\Gamma$ and $\bvec q\approx K^{(\prime)}$. The $K$-valley (left subpanel) includes the Fermi contour.}
    \label{fig:magnetism}
\end{figure}

\paragraph*{Spin fluctuations \& magnetic instabilities.}
To study magnetic correlations in few-layer graphene we use a random phase approximation (RPA) approach. We calculate the free electronic susceptibility $\hat\chi_0(q) = \Tr_k[\hat G_0^{\vphantom{T}}(k-q)\odot\hat G_0^T(k)]$ in the static limit $iq_0\to0$, where $\hat G_0(k)$ is the free Matsubara Green's function as a matrix in orbital space, $k=(ik_0,\bvec k)$ the electronic ``four-momentum'', and ``$\odot$'' an element-wise matrix product. Assuming a local Hubbard interaction $U$ allows us to resum the infinite ladder series in the crossed particle-hole channel to arrive at a Stoner-like criterion~\cite{klebl2019inherited} for the orbital-resolved susceptibility $\hat\chi_0(\bvec q)$: The critical interaction strength required for the onset of magnetic order is given by $U_\mathrm{c} = 1/\chi_0^m(\bvec q^m)$, with $\chi_0^m(\bvec q^m)$ the largest eigenvalue of $\hat\chi_0(\bvec q^m)$ and $\bvec q^m$ the transfer momentum at which the maximum occurs. We adapt $T =5\cdot10^{-5}$ eV as temperature broadening for $\hat\chi_0(\bvec q)$ throughout this work, see SM~\cite{SM} for more details on the orbital-space RPA.

\Cref{fig:magnetism}~(a) demonstrates that magnetism is potentially realized in ABCB tetralayer graphene near the valence band VHS. The critical interaction strength drops to $U_\mathrm{c}\approx4.5\,\mathrm{eV}$, which is of similar order as estimates of the effective (screened) on-site $U$ in graphene~\cite{wehling2011strength, rosner2015wannier}. In \cref{fig:magnetism}~(b) we observe that the valence band VHS drives ferrimagnetic fluctuations within the C-layer of ABCB graphene, while away from the VHS layer-agnostic antiferromagnetic fluctuations gain in importance. The momentum structure of the magnetic susceptibility is shown in panel~(c), where we plot the leading susceptibility eigenvalue for all $\bvec q$ close to $\Gamma$ and around the $K^{(\prime)}$ points --- these are the areas in momentum space where we expect $\chi^m_0(\bvec q)$ to pick up structure by either the VHSs or nesting. While the detailed weight distribution in $\chi^m_0(\bvec q)$ evolves upon doping across the valence band VHS, we observe that the leading contribution always stems from momenta close to $\Gamma$. Therefore intravalley scatterings and the divergent DOS mainly drive magnetic order. Other orders as, e.g., intervalley coherent order, may emerge as subsequent instabilities from the magnetic one in the presence of purely local interactions. For the VHS on the electron-doped side, a similar picture of magnetic instabilities emerges (see SM~\cite{SM}), allowing us to focus on the hole-doped side.

\paragraph*{Pairing from local interactions.}
\begin{figure}
    \centering
    \includegraphics[width=0.5\textwidth]{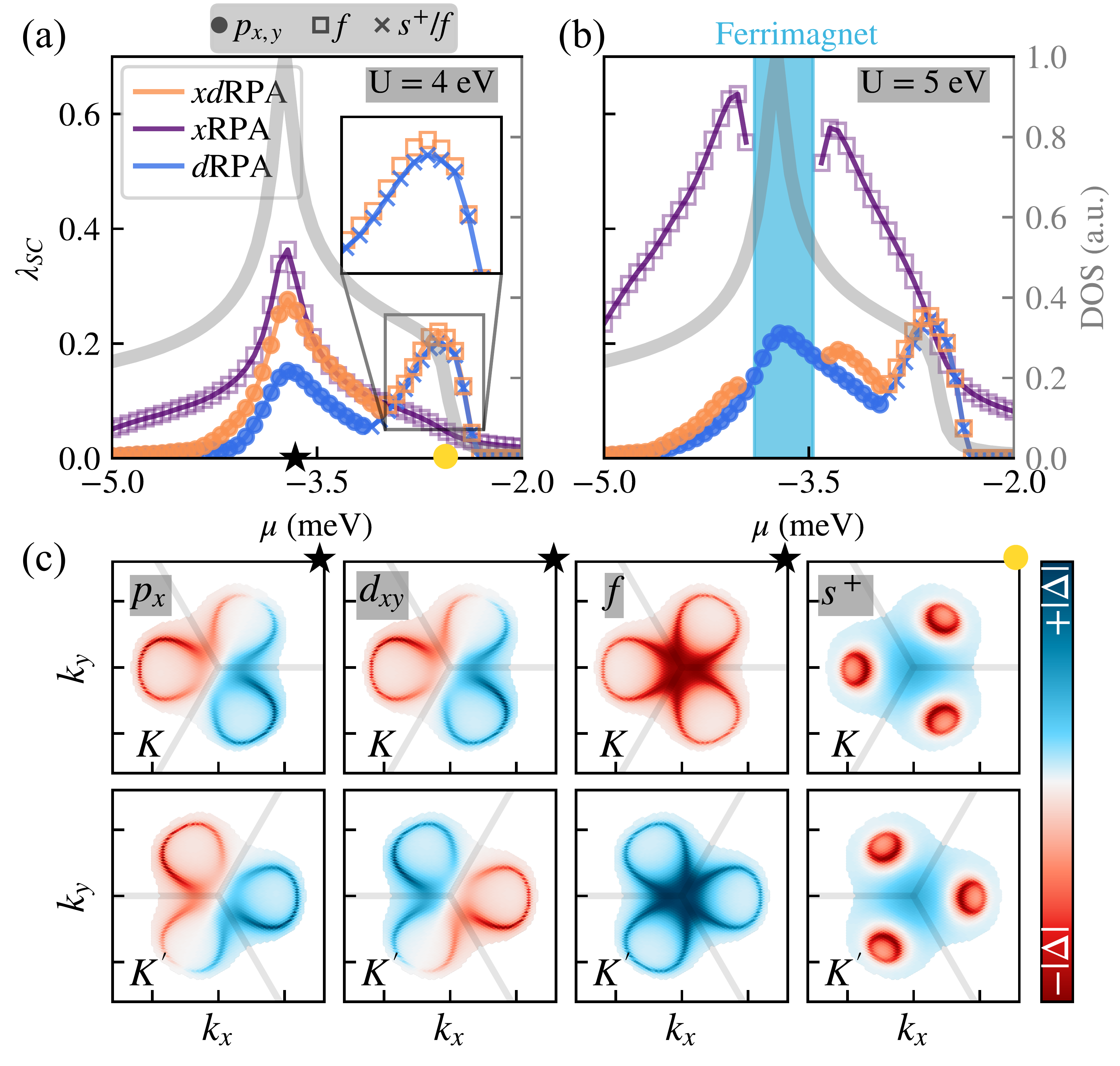}
    \caption{
    Superconducting instabilities mediated by spin \&charge fluctuations in ABCB graphene. Panels~(a,b) show the leading SC coupling constant $\lambda_{SC}$ as function of the chemical potential $\mu$. Local Hubbard-$U$ interactions particularly support spin-fluctuations, which generically enhance spin-triplet $f$-wave superconductivity due to the allowed intervalley scattering (\xRPA{}). Screening from long-ranged Coulomb interactions (\dRPA{}) instead favors doubly-degenerate $p_{x,y}$-wave superconducting order near the VHS and almost degenerate $f/s^+$-wave SC order in regions of the three-pocket Fermi surface near the valence band edge. Combining the effect of spin \&charge fluctuations (\xdRPA{}), we find that local interactions are screened significantly by the long-ranged Ohno interaction and that SC order similar to the screening scenario (\dRPA{}) prevails and is further amplified by spin-fluctuation near the Stoner instability. For $U=5\,$eV, (b) the Stoner instability suppresses superconductivity from spin-fluctuations due to the presence of ferromagnetic ordering and only screening driven SC order survives. The corresponding instabilities in the valleys $K$ and $K^{\prime}$ are displayed in panel (c).}
\label{fig:fig3}
\end{figure}

As established in the previous paragraph, local interactions cause ferrimagnetic correlations to prevail in ABCB graphene on the hole-doped side. This is particularly striking as large areas in the phase diagram of bilayer graphene are governed by symmetry-broken phases that show clear indications of hysteresis~\cite{seiler2022quantum}. Close to the ferrimagnetic instability, spin fluctuations arising from local, repulsive interactions can provide the pairing glue for an unconventional superconducting state~\cite{scalapino2012common,graser2009near,fischer2022unconventional,fischer2021spin,stauber2019,gonzalez2023universal}. The effective pairing vertex in static approximation for this Kohn-Luttinger like mechanism includes contributions from local interactions to the exchange- and direct particle-hole channels~\cite{graser2009near}
\begin{multline}
    V^{\text{PP}}_X(\bvec k, \bvec k') = \hat U - \frac{\hat{U}\hat{\chi}_{0}(\bvec q_X) \hat U }{\mathbb{1} - \hat U\hat{\chi}_{0}(\bvec q_X)} \delta_{\bvec q_X, \bvec k + \bvec k'} + {}\\
    \frac{ \left [  \hat{U} \hat{\chi}_{0}(\bvec q_D) \right]^2 \hat U }{\mathbb{1} - \left [ \hat{U} \hat{\chi}_{0}(\bvec q_D) \right ]^2} \delta_{\bvec q_D, \bvec k - \bvec k'},   
\end{multline}
where $U$ is the local onsite interaction, $\bvec q_{X(D)}$ denotes the dominant momentum transfer in the exchange (direct) particle-hole channel and the hat symbol denotes the orbital structure of interactions and polarization function, see SM~\cite{SM} for details. Within the spin fluctuation exchange formalism (\xRPA{}), we finally solve the linearized gap equation in the projected subspace containing the bands in an energy window $W$ around the Fermi level
\begin{equation}
    \lambda_{\text{SC}}  \Delta_{b}(\bvec k) = - \frac{1}{N_{\bvec k}}  \sum_{\bvec k' b'} \hat{V}^{\text{PP}}_{S, bb'}(\bvec k, \bvec k') \chi^{\text{PP}}_{b'}(\bvec k') \Delta_{b'}(\bvec k'),
\end{equation}
where $\hat V^\mathrm{PP}_{S, bb'}$ and $\chi^\mathrm{PP}_{b'}$ denote the band-projected effective pairing vertex and particle-particle susceptibility, respectively. The largest coupling constant therefore gives an estimate for the critical temperature $T_c = T\,e^{-1/\lambda_{SC}}$ of the superconducting (SC) transition~\cite{raghu2010superconductivity} and the corresponding eigenfunction yields the symmetry of the SC order parameter $\Delta_b(\bvec k)$.

Since local interactions are flat in momentum space, they induce both inter- and intravalley coupling. In conjunction with single layer ferromagnetic fluctuations, the intervalley exchange enhances order parameters that are (i) spin-triplet and (ii) change sign under a valley flip. We find that local interactions exclusively promote valley-singlet, spin-triplet $f$-wave order for all fillings around the VHS in ABCB stacked graphene. The coupling constant $\lambda_{SC}$ is enhanced close to the Stoner transition for $U=4$ eV as plotted in purple in Fig.~\ref{fig:fig3}~(a), while it decreases continuously when doping away from the VHS. As soon as the local interaction strength $U$ exceeds the critical value of the Stoner transition $U_c$, superconductivity is suppressed in the Stoner regime and the peak of the coupling constant $\lambda_{SC}$ splits into two as shown in Fig.~\ref{fig:fig3}~(b). We note that further screening of local interactions may therefore shift the position of superconducting regions observed in experiment~\cite{stepanov2020untying}.

\paragraph*{Pairing from screened Coulomb interactions.}
Screened Coulomb repulsion was discussed to provide the dominant pairing glue in bi- and trilayer graphene~\cite{ghazaryan2022annular,ghazaryan2023multilayer,li2023charge,cea2022superconductivity,jimeno2023superconductivity,pantaleon2023superconductivity}. In the continuum theories employed by a majority of works, the additional flavor degeneracy due to the valley degree of freedom enhances the polarization function $\chi_0(\bvec q)$ by a factor of $4$ instead of $2$ compared to the usual spin $SU(2)$, which boosts the contribution of charge-fluctuations to electron mediated pairing. However, effective continuum theories hinder a direct description of the microscopic interactions containing long- and short-ranged terms. This manifests in an artificial spin/valley $SU(4)$ symmetry that must be lifted by an phenomenological inter-valley Hunds coupling~\cite{ghazaryan2022annular,ghazaryan2023multilayer}. Here, we remedy this shortcoming by exploiting the orbital-resolved RPA approach presented in this manuscript and use the ``Ohno'' interaction profile~\cite{ohno1964some,wehling2011strength}
\begin{equation}
    V^{O}(\bvec r) = \frac{Ua}{\sqrt{a^2+\bvec r^2}}e^{-|\bvec r|/d},
\label{eq:ohno_interaction}
\end{equation}
with realistic parameters $a = 0.3 a_0$, $d = 200 a_0$ and $a_0 = 2.46\,\text{\AA}$ the graphene lattice constant. The parameter $d$ controls the external screening from, e.g., metallic gates or the substrate, $a$ controls the atomic scale decay of the Coulomb interaction in real space, and $U$ sets the on-site (Hubbard) interaction strength. Pairing due to interaction-induced screening is then captured by the effective vertex
\begin{equation}
    \hat V^{\text{PP}}_D(\bvec k, \bvec k') = \frac{\hat{V}^{O}(\bvec{q}_D)}{\mathbb{1} + 2 \hat{V}^{O}(\bvec{q}_D)  \hat{\chi}_{0}(\bvec{q}_D)} \delta_{\bvec{q}_D, \bvec k - \bvec k'},
\label{eq:vertex_d}
\end{equation}
where $\hat{V}^{O}(\bvec q)$ is the discrete Fourier transform of the Ohno interaction matrix obtained from Eq.~\eqref{eq:ohno_interaction} and $\bvec{q}_D$ is the relevant momentum transfer in the direct particle-hole channel. Taking only screened Coulomb interactions into account (\dRPA{}), the superconducting coupling constant shows two peaks that are located near the VHS and the edge of the valence bands, see Fig.~\ref{fig:fig3}~(a,b). 

Near the VHS, the superconducting order parameter with highest transition temperature is a doubly degenerate spin-triplet with $p_{x,y}$ symmetry that changes sign within each valley as shown in panel (c).

We argue that order parameters with an intravalley sign change are generically favored by the screened Coulomb interaction for realistic interaction parameters: Microscopically, screening induced pairing Eq.~\eqref{eq:vertex_d} is most relevant for momenta $\bvec{q}_D$ where the polarization function $\hat{\chi}_0(\bvec{q}_D)$ peaks.
\Cref{fig:magnetism}~(c) demonstrates that these momenta are $\bvec q_D\approx\Gamma$. Therefore the repulsive Coulomb interaction $\hat V^O(\bvec q_D)$ gets suppressed significantly close to $\Gamma$. Nevertheless the screened interaction $\hat V^{\text{PP}}_D(\bvec k, \bvec k')$ is still mostly positive in momentum space, c.f. Eq.~\eqref{eq:vertex_d} and SM~\cite{SM}, such that possible order parameters must change sign on the respective Fermi surface sheets connected by the vector $\bvec{q}_D$. Scattering of Cooper pairs is therefore selectively enhanced within the same valley ($\bvec k \approx \bvec k'$). On the other hand, intervalley scattering $\bvec{q}_D \approx K^{(\prime)}$ caused by short-ranged terms in the Coulomb interaction is suppressed. We note that this competition of inter- vs. intravalley scattering can only be resolved when considering the full Coulomb interaction with respective ratios of on-site and long range interaction strengths. 

\begin{figure}
    \centering
    \includegraphics[width=0.48\textwidth]{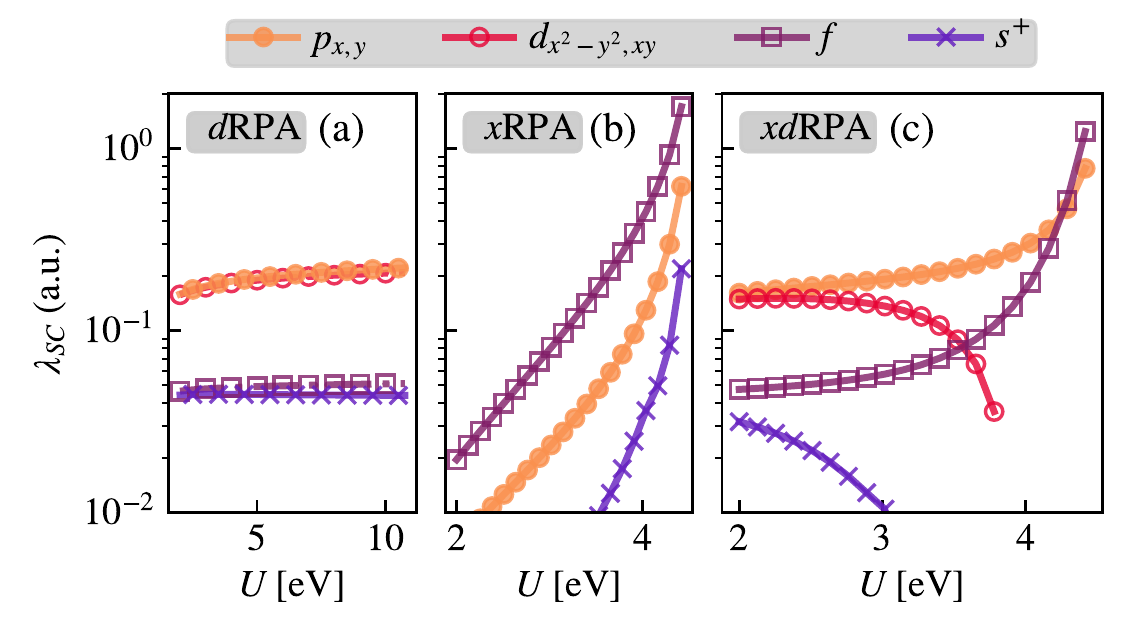}
    \caption{Competition of superconducting order at the VHS of ABCB for realistic values of the on-site interaction $U$ separated into contribution from spin-fluctuations and screening. (a) Screening of long-ranged Coulomb interactions (\dRPA{}) generically favors $p_{x,y}$ ($d_{x^2-y^2,xy}$)-wave order parameters that change sign within the same valley, while $f/s^+$ order is suppressed for all values of $U$. (b) Spin-fluctuation exchange (\xRPA{}) is dominated by local interactions and favors spin-triplet $f$-wave order. (c) Combining screening and spin-fluctuations (\xdRPA{}) reveals a delicate interplay of above mechanisms. The degeneracy between $p/d$-wave is lifted as spin-singlet order is suppressed by spin-fluctuations. Instead, $f$-wave order is enhanced and eventually exceeds $p$-wave order at the Stoner transition.}
    \label{fig:fig4}
\end{figure}

Indeed the two spin-triplet $p_{x,y}$-wave order parameters are accompanied by two spin-singlet $d_{x^2-y^2,xy}$ order parameters. The latter come with a slightly lower coupling constant $\lambda_{SC}$ at the VHS, see Fig.~\ref{fig:fig4}.
\Cref{fig:fig3}~(a,b) further demonstrates that within the \dRPA{} approach, $f$-wave and $s^+$-wave symmetric gap functions are suppressed around the VHS for realistic values of on-site $U$ such that the $p_{x,y}$-wave pairing prevails. Varying $d$ has no influence on the \dRPA{} results, see SM~\cite{SM}.

The situation at the valence band edge is fundamentally different: Here, the increase of the superconducting coupling constant $\lambda_{SC}$ can be traced back to the trigonal warping related three pocket structure of the Fermi surface in ABCB. As the Fermi surface shrinks, $p_{x,y}$-wave order is suppressed because an intravalley sign change is increasingly disfavored.
Instead almost degenerate $f$- and $s^+$-wave order prevails, which is driven by local interactions as in the \xRPA{} (spin) case. The critical temperature reaches values up to $T_c \approx 10$ mK and the predominant $f/s^+$-wave order stays unmodified for different (numerical) momentum resolutions, see SM~\cite{SM}.


\paragraph*{Competition of spin \& charge fluctuations.}
Finally, we study the mutual interference of the charge and spin mechanisms discussed above by carrying out independent resummations of (i) short range interactions in the exchange channel and (ii) long range interactions in the direct channel (\xdRPA{}), see SM~\cite{SM}. We observe that values of the coupling constant $\lambda_{SC}$ lie between those of the spin fluctuation (\xRPA{}) and the screening mechanism (\dRPA{}) with critical temperatures reaching $T_c \approx 25$ mK, whereas the order parameter symmetry shows $p_{x,y}$-wave order near the VHS and $f$-wave order at the valence band edge. Even though the overall density dependence and symmetry of the superconducting state resembles pure screening driven SC, important insights can be gained from this analysis.  First, we note that the coupling constant $\lambda_{SC}$ is enhanced compared to the pure screening mechanism as shown in \cref{fig:fig3}. Therefore, we conclude that spin-fluctuations support charge-fluctuations in the formation of superconductivity in ABCB, especially close to the Stoner transition. 
Second, the approximate degeneracies between $f$- and $s^+$-wave order close to the valence band edge are significantly lifted in the combined \xdRPA{} approach, as well as between $p_{x,y}$- and $d_{x^2-y^2,xy}$-wave order around the VHS. \Cref{fig:fig4} presents a refined analysis of the superconducting tendencies for $\mu=\mu_\mathrm{VHS}$ for different values of the on-site interaction strength $U$. Since fluctuations around the ferrimagnetic phase suppress spin-singlet superconductivity, the nearly degenerate $f/s^+$- and $p_{x,y}/d_{x^2-y^2,xy}$-wave order parameters acquire significant splitting with increasing $U$ and only the triplet order parameters ($f$- and $p_{x,y}$-wave) prevail.
Third, the pairing glue arising from local interactions is significantly reduced by the long range tail of the Coulomb repulsion such that spin fluctuation induced pairing becomes effectively suppressed far from the Stoner transition. This is underlined by the observation that for larger on-site interactions [$U=5\,\mathrm{eV}$, cf. \cref{fig:fig3}~(b)], spin fluctuation driven SC (purple line) is visibly enhanced due to the vicinity to the magnetic state, while the effective coupling constant $\lambda_{SC}$ within the combined \xdRPA{} approach (orange line) is merely affected. At the VHS the situation is more delicate: As $U\to U_\mathrm{c}$, we see the mutual enhancement of $p_{x,y}$- and $f$-wave order parameters which benefit from the spin-fluctuation enhancement, see \cref{fig:fig4}. In the immediate vicinity to the magnetic instability the \xRPA{} behavior is recovered such that for $U\lesssim U_\mathrm{c}$ the leading instability is again of $f$-wave character. While our combined \xdRPA{} approach performs resummations in both particle-hole interaction channels (crossed \& direct), it cannot capture inter-channel feedback, which could account for screened local interactions at the VHS, i.e., suppress the magnetic instability, and therefore boost $f$-wave SC at the VHS.

\paragraph*{Discussion.}

In this work we discuss the role of spin fluctuations and screening on the formation of correlated magnetic and superconducting states in tetralayer graphene with emphasis on ABCB stacking. The particular shape and orbital polarization of the bands in ABCB on the hole-doped side resembles the bandstructure of bilayer graphene in an external electric field. Since ABCB graphene is not topologically obstructed as relative twisted graphene stacks, we expect our result to be qualitatively transferable to other stacking configurations of few-layer graphene, such as AB bilayer graphene. In contrast to ABCB, superconductivity in AB graphene emerges at relatively high (experimentally accessible) displacement 
fields~\cite{zhou2021superconductivity,zhou2022isospin,zhang2023enhanced,holleis2023ising}, implying that ABCB graphene provides an interesting avenue to further stabilize superconductivity in graphene-based multilayers. Due to the metastable nature of ABCB graphene \cite{Wirth2022Oct}, the experimental realization of transport devices is likely to pose special challenges for the sample preparation, especially for the encapsulation of ABCB graphene in hexagonal boron nitride like it is known from ABC trilayer graphene \cite{zhou2021half, Yang2019Dec}.

Our results further raise important consequences originating from the orbital-resolved weak coupling treatment that captures long and short-ranged Coulomb interactions consistently. First, we argue that for realistic values of the on-site interaction~\cite{wehling2011strength, rosner2015wannier} $p_{x,y}$-wave SC order will always prevail within a pure screening mechanism. This is in contrast to previous works~\cite{jimeno2023superconductivity,li2023charge} that show emerging $f/s^+$-wave superconductivity at the VHS in bi- and tetralayer graphene stacks resulting from screened Coulomb interactions alone. Second, we observe another peak in the superconducting coupling constant $\lambda_{SC}$ for low hole-doping with emergent $f/s^+$-wave symmetry. Previous works by Ghazaryan \textit{et.~al}~\cite{ghazaryan2022annular, ghazaryan2023multilayer} report a slim peak near the valence band edge with dominant $d$-wave symmetry, which contradicts our results and can be traced back to the effect of local interactions that are captured within the orbital-resolved RPA. Our \xdRPA{} approach further highlights the importance of studying electronic phases in multilayer graphene stacks with methods that (i)~allow for inter-channel feedback, e.g. the functional renormalization group~\cite{metzner2012functional,qin2023functional},  and (ii)~incorporate local Coulomb repulsion on the same footing as the long range tail.

\paragraph*{Acknowledgments.}
We thank C. Stampfer, T. Taubner and K. Wirth for fruitful discussions. This work was supported by the Excellence Initiative of the German federal and state governments, the Ministry of Innovation of North Rhine-Westphalia and the Deutsche Forschungsgemeinschaft (DFG, German Research Foundation). 
JBH, LK, AF and DMK acknowledge funding by the DFG under RTG 1995, within the Priority Program SPP 2244 ``2DMP'' --- 443273985.
LK and TOW greatfully acknowledge support from the DFG through FOR 5249 (QUAST, Project No. 449872909) and SPP 2244 (Project No. 422707584).
TOW is supported by the Cluster of Excellence ``CUI: Advanced Imaging of Matter'' of the DFG (EXC 2056, Project ID 390715994).
DMK acknowledges support by the Max Planck-New York City Center for Nonequilibrium Quantum Phenomena. We acknowledge computational resources provided by the Max Planck Computing and Data Facility, RWTH Aachen University under project numbers rwth0742 and rwth0716, and through the JARA Vergabegremium on the JARA Partition part of the supercomputer JURECA~\cite{JURECA} at Forschungszentrum Jülich.
\bibliography{literature4LG}

\supplement{Supplemental Material: \\ Spin and Charge Fluctuation Induced Pairing in ABCB Tetralayer Graphene}

\section{Non-interacting Hamiltonian}
Our tight binding Hamiltonian for graphene tetralayers is a modified Slonzecewski-Weiss-McClure (SWMC) Hamiltonian, with parameters chosen based on Ref.~\cite{Rubio_flg_Tb} summarized in Table~\ref{tab:TBparams}. The Hamilton matrix can be constructed as
\begin{equation}
    H_{i,j}^{y}(\bvec k) = \delta_{i,j}\delta_{i,A} \Delta_{y}
    +\sum_{u_1,u_2 \in \mathbb{Z}} \sum_n \,\delta_{d_{i,j}^{u_1,u_2},d_n}e^{-i\bvec k \bvec R(u_1,u_2)}\gamma_n \,,
\end{equation}
where $i$ and $j$ are site indices within the unit cell, $y$ marks the type of the lattice and is either ABAB, ABCA or ABCB and $\delta_{i,A}$ is one if $i$ is an a-site and $0$ otherwise. $u_1$ and $u_2$ iterate over all unit cells in the infinite lattice, $\bvec R(u_1,u_2)$ gives the vectorial distance between the unit cells and $d_{i,j}^{u_1,u_2}$ gives the distance between site $i$ and the image of site $j$ shifted by $u_1$ and $u_2$ unit-cell vectors.
\begin{table}[!tbh]
    \centering
    \caption{Slonzecewski-Weiss-McClure Hamiltonian model parameters.}\label{tab:TBparams}
    \vspace{0.5em}
    \begin{ruledtabular}
    \def\arraystretch{1.5}
    \begin{tabular}{lll}
        Name & Value in eV & Distance in \AA \\
        \hline
        $\gamma_0$ & $\phantom{-}2.553$ & $1.42$\\
        $\gamma_1$ & $\phantom{-}0.343$ & $3.35$\\
        $\gamma_2$ & $-0.009$ & $6.70$ \\
        $\gamma_3$ & $\phantom{-}0.18$ &  $4.16$\\
        $\gamma_4$ & $\phantom{-}0.173$ & $3.64$\\
        $\gamma_5$ & $\phantom{-}0.018$ & $6.85$\\
        $\Delta_{\text{ABAB}}$ &  $-0.003$ & $0.0$, a site\\ 
        $\Delta_{\text{ABCA}} $&  $\phantom{-}0.0$& $0.0$, a site\\
        $\Delta_{\text{ABCB}} $& $-0.018$ & $0.0$, a site
    \end{tabular}
    \end{ruledtabular}
\end{table}

\section{Random Phase Approximation}
To calculate the non-interacting susceptibility $\hat\chi_0(\bvec q)$ in static approximation, we must carry out the summation
\begin{equation}
\begin{split}
    \chi_0^{oo'}(\bvec q) &= -\bigg( \lim_{iq_0\to0} \Tr_{k}\big[ \hat G(k-q) \odot \hat G(k)^T \big] \bigg)_{oo'} \\
    &= -\frac1{N_{\bvec k}} \sum_{\bvec k, b, b'} \frac{n_F(\epsilon_b(\bvec k))-n_F(\epsilon_{b'}(\bvec k-\bvec q))}{\epsilon_b(\bvec k) - \epsilon_{b'}(\bvec k-\bvec q)} \, u_{ob}(\bvec k) u_{o'b}^*(\bvec k) u_{ob'}^*(\bvec k-\bvec q) u_{o'b'}(\bvec k-\bvec q) \,,
\end{split}
\label{eq:supp-loop}
\end{equation}
where $\epsilon_b(\bvec k)$ denotes the dispersion relation and $u_{ob}(\bvec k)$ the Bloch function at momentum $\bvec k$ for band $b$ and orbital (site) $o$.
In the actual numerical implementation, we make use of the fact that the only momenta $\bvec q$ relevant for magnetic instabilities are  those where $\bvec q\approx \bvec k_F \pm \bvec k'_F$, with $\bvec k_F^{(\prime)}$ on the Fermi surface. We thus choose an energy widow $W\gg T$ (in our calculation $W=3\,\mathrm{meV}$ and $T=0.025\dots0.1\,\mathrm{meV}$) and obtain all momenta $\bvec q \in \{ \bvec k \pm \bvec k' \text{ with } \bvec k,\bvec k' \in \mathbb{FS}_W \}$ for the extended Fermi surface $\mathbb{FS}_W = \{ \bvec k \in \mathbb{BZ} \text{ s.t. } \exists b \text{ with } |\epsilon_b(\bvec k) - \mu| < W \}$.
The calculation of $\hat\chi_0(\bvec q)$ is perfectly parallel in $\bvec q$. Moreover, we parallelize the summation over $\bvec k$ to obtain higher performance in our custom CUDA kernels. We additionally note that we make use of a cached dispersion relation and obrital to band transform on a fixed momentum mesh (we checked $4800\times4800$ and $7200\times7200$ points).

From the free electronic susceptibility $\hat\chi_0(\bvec q)$ on the relevant $\bvec q$, we calculate the critical on-site interaction strength required for the onset of magnetic order. The crossed particle-hole channel RPA resummation of the effective interaction yields
\begin{equation}
    \label{eq:rpavertex}
    \hat W(\bvec q) = \frac{\hat U}{\mathbb{1}-\hat U\hat\chi_0(\bvec q)} \,,
\end{equation}
where we directly observe that for an on-site Hubbard interaction $\hat U=U\mathbb 1$, the maximum eigenvalue of $\hat\chi_0(\bvec q)$ over all $\bvec q$ determines (i) the critical interaction strength $U_\mathrm{c}=1/\chi^m_0(\bvec q^m)$ and (ii) the type of magnetic order via the corresponding eigenvector $\vec{\mathcal{X}}^m_0(\bvec q^m)$.

\subsection{Flowing random phase approximation}
\label{sec:flow-rpa}
An alternative way to look at the RPA from above is via a single-channel functional renormalization group (FRG) flow. From the diagrams contained in the two particle irreducible flow with neglected self-energies~\cite{beyer2022reference} it is evident that the effective FRG vertex in static approximation sums up all diagrams contained in the effective RPA vertex \cref{eq:rpavertex} when restricted to the crossed particle hole channel. We therefore rewrite the RPA as a single channel FRG, i.e.,
\begin{equation}
    \dot{\hat W}_\Lambda(\bvec q) = -\hat W_\Lambda(\bvec q) \dot{\hat L}_\Lambda(\bvec q) \hat W_\Lambda(\bvec q) \,.
\end{equation}
One quickly verifies that the solution to this differential equation is given by
\begin{equation}
    \hat W_\Lambda(\bvec q) = \frac{\hat W_\infty(\bvec q)}{\mathbb{1}-\hat W_\infty(\bvec q)\int_\Lambda^\infty\mathrm{d}\Lambda'\dot{\hat L}_{\Lambda'}(\bvec q)} \,.
\end{equation}
From the FRG perspective, it is therefore sufficient to calculate the particle-hole loop integral to get an expression for the free susceptibility at scale $T$:
\begin{equation}
    \hat\chi^\mathrm{FRG}_T(\bvec q) = \int_T^\infty\mathrm{d}\Lambda \,\dot{\hat L}_\Lambda(\bvec q) = \int_T^\infty\frac{\mathrm{d}\Lambda}{2\pi} \bigg[ \frac1{N_{\bvec k}}\sum_{\bvec k} \hat G(i\Lambda, \bvec k) \odot \hat G^T(i\Lambda, \bvec k - \bvec q) + \mathrm{h.c.} \bigg] \,.
\end{equation}
Note that we employ a sharp frequency cutoff here. $\hat\chi^\mathrm{FRG}_T(\bvec q)$ resembles an approximation to the Matsubara summation in the explicit RPA, where the frequencies above the lowest one are treated as continuum. Such an approximation is similar to the explicit Matsubara summation used in large unit cell systems in Ref.~\cite{klebl2019inherited}. We observe that in the flowing RPA formulation, though, we can make use of fast Fourier transforms (FFTs) to calculate the convolution over $\bvec k$ and $\bvec q$. For large momentum meshes, the numerical scaling of this operation is $\mathcal O(N_{\bvec k}\log N_{\bvec k})$ instead of $\mathcal O(N_{\bvec k}N_{\bvec q})$ for the explicit sum over $\bvec k$ for each $\bvec q$. Using the flowing RPA trick with FFTs allows us to bump the momentum resolutions significantly higher than using the explicit RPA: We calculated the flowing RPA susceptibilities for a momentum resolution of $N_{\bvec k} = 12288^2 = (2^{12}\cdot 3)^2$ points and confirmed that indeed our calculations are converged. Note that numerically, the flowing RPA covering roughly 400 adaptively chosen points in the $\Lambda$ integral were \emph{significantly faster} on a single CPU node of the JURECA cluster~\cite{JURECA} than the explicit RPA CUDA kernels on four NVIDIA A100 GPUs with approximately half the resolution ($N_{\bvec k}=7200^2$).

\subsection{Magnetic instabilities for other parameters}
\begin{figure}
    \centering
    \includegraphics{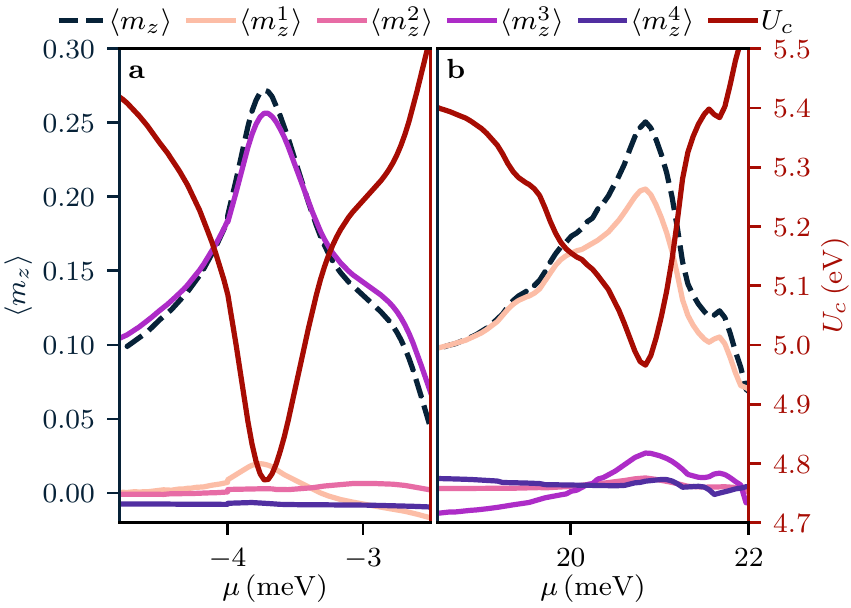}
    \caption{Critical interaction strength (red) and layer magnetization curves for the hole-doped~(a) and electron-doped~(b) van Hove singularity in ABCB graphene. In contrast to the hole-doped side, the electron-doped carries most magnetization in layer A~(1). Apart from that, the qualitative features are equivalent. The dominant ferrimagnetism in layer~1~[panel (a)] corresponds to the layer polarization of the Bloch functions being mostly in layer~1 (cf. \cref{fig:bands}).}
    \label{fig:magnetism-other-side}
\end{figure}

To confirm that the conduction band physics is qualitatively equivalent the effects in the valence band, we show the magnetic instabilities for both sides at $T=10^{-4}\,\mathrm{eV}$ in \cref{fig:magnetism-other-side}.

\begin{figure}
    \centering
    \includegraphics{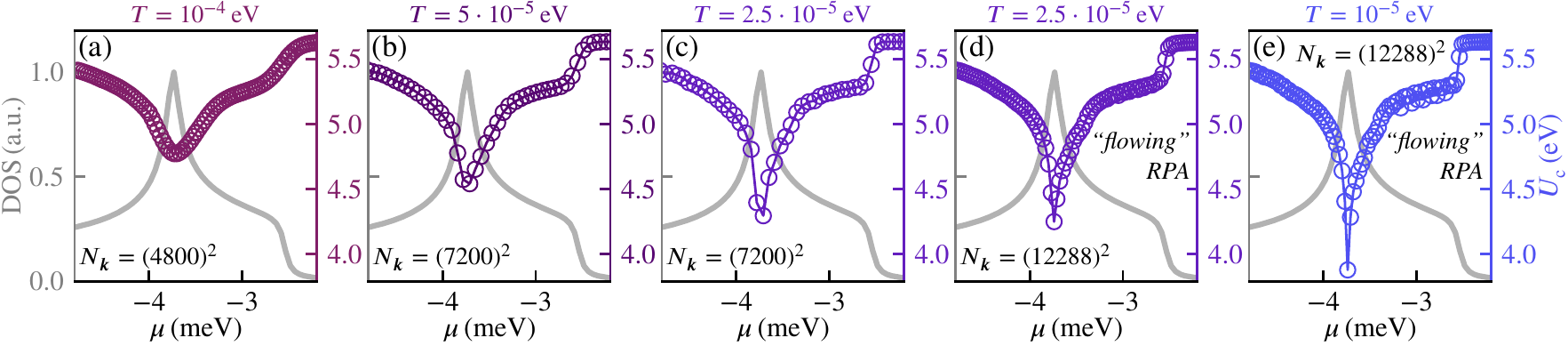}
    \caption{Critical interaction strength as a function of filling for a range of temperatures and momentum resolutions. Panels~(a-c) use the explicit RPA summations and panels~(d,e) the flowing RPA.}
    \label{fig:chi-temperatures}
\end{figure}

The behavior of $\hat\chi_0(\bvec q)$ as a function of temperature is depicted in \cref{fig:chi-temperatures}. Note that we include high resolution, low temperature calculations that were obtained with the \emph{flowing RPA} trick presented in \cref{sec:flow-rpa} in panels~(d,e).
\FloatBarrier

\begin{figure}%
    \centering%
    \includegraphics{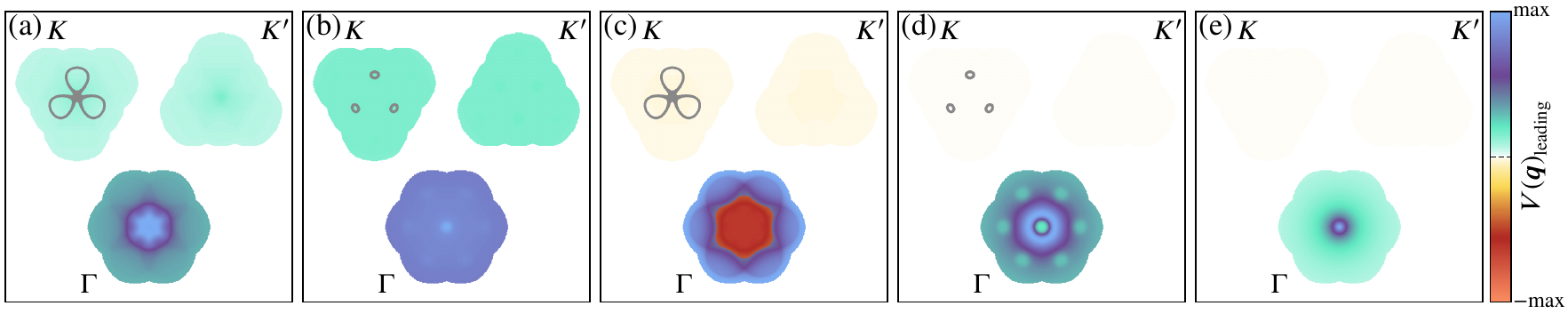}
    \includegraphics{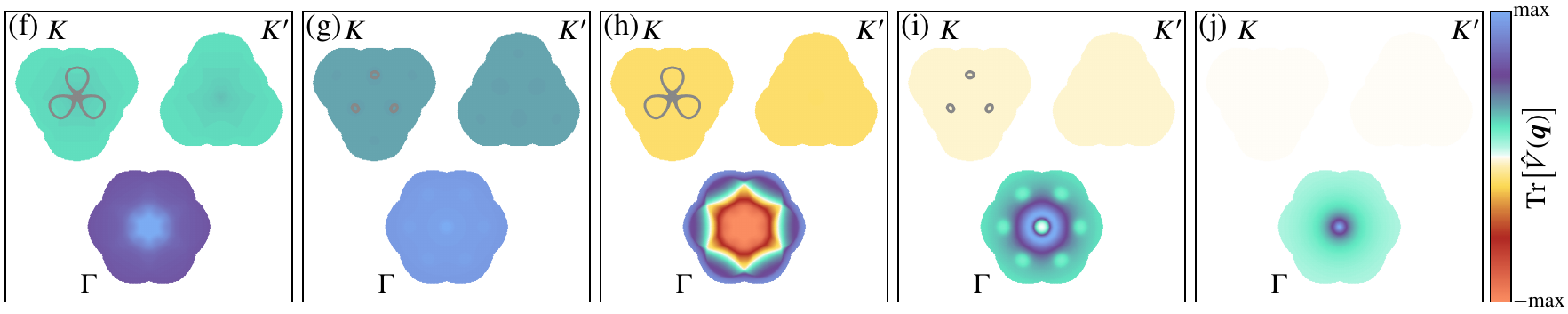}
    \includegraphics{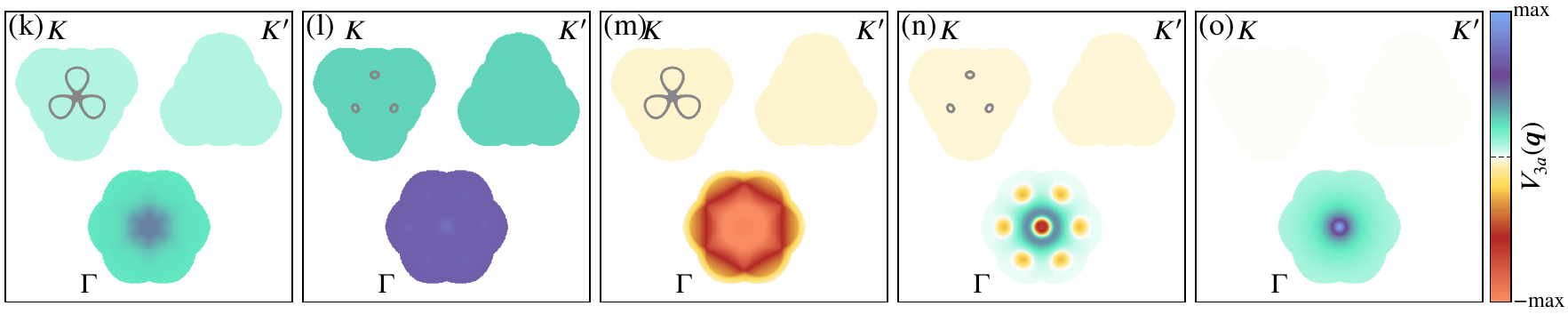}
    \includegraphics{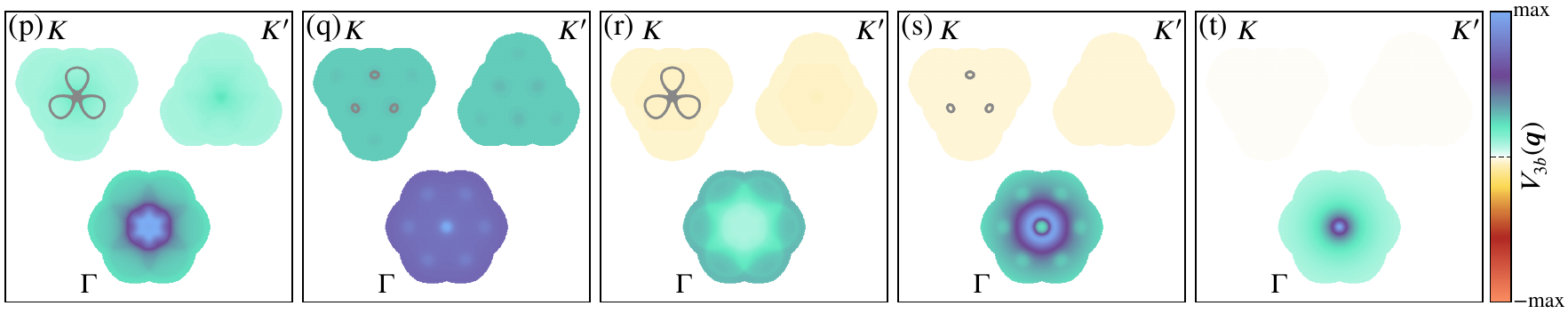}
    \caption{Momentum and orbital structure of effective vertices. First column: \xRPA{} vertex for $\mu\approx\mu_\mathrm{VHS}$, second column: \xRPA{} vertex for $\mu$ at the valence band edge, third column: \dRPA{} vertex for $\mu\approx\mu_\mathrm{VHS}$, fourth column: \dRPA{} vertex for $\mu$ at the valence band edge, and fifth column: bare Ohno vertex. We display the leading vertex eigenvalue in the first row~(a-e), the trace over the orbitals in the second row~(f-j), the projection of the vertex to the a-site of the C layer (layer 3) in the third row~(k-o), and the projection to the b-site of the C layer (layer 3) in the fourth row~(p-t). The colorbar in each panel is normalized individually, with panels~(k,p), (l,q), (m,r), (n,s), and~(o,t) each sharing the normalization. In the orbital projections we observe that attraction is generated only on the a-site of layer 3, such that it is projected out by the particle-particle loop in the linearized gap equation~\cref{eq:lin_gap} and the effective vertex relevant for superconductivity remains repulsive in momentum space. In the subpanels around the $K$ point, we additionally include the Fermi contour for scale.}
    \label{fig:vertices}
\end{figure}

\section{Superconductivity}
To study unconventional superconductivity driven by electronic interactions, we study different mechanisms that are based on (i) spin-fluctuations by local interactions ($x$RPA) (ii) charge-fluctuations due to screening of the long-ranged Coulomb interactions ($xd$RPA) and (iii) a combined approach ($d$RPA) that captures a superset of diagrams contained in the mechanisms above. This section is denoted to show the corresponding diagrams describing the contributions of either mechanism to superconducting pairing and discuss their practical implementation.

\subsection{Local Interactions}
In the vicinity of magnetic instabilities, spin and charge fluctuations that are driven by local Hubbard-$U$ interactions can provide the pairing glue for an unconventional superconducting state. To capture the effect of spin-fluctuation induced pairing in ABCB, we resort to the well-known diagrammatic expansion~\cite{graser2009near,scalapino1987} based on transversal and longitudinal spin-fluctuations as shown in \Cref{fig:feynman-diagrams}. Even though orbital degrees of freedoms are not explicitly shown in the diagrams, we keep the full orbital dependence of the susceptibility and the respective interactions throughout this work. Within this approach we capture all diagrams from the exchange channel (transversal spin-fluctuations) and the subgroup of diagrams in the direct particle-hole channel that contain an even number of loops (longitudinal spin-fluctuations). As we assume the system to be in the paramagnetic phase, i.e. the normal-state Hamiltonian is $SU(2)$ symmetric, the susceptibilities carry no explicit spin degree of freedom and it is possible to only consider the pairing vertex, where the ingoing- and outgoing spins are opposite. All spin-singlet (spin-triplet) instabilities are then captured in the anti-symmetric (symmetric) sector of the pairing vertex. The transversal and longitudinal spin-fluctuation diagrams translate to 
\begin{align}
\begin{split}
    \left [ V^{\text{PP}}_X \right ]_{o_1,o_2,o_3,o_4}(\bvec k, \bvec k^{\prime}) = U \delta_{o_1,o_2}\delta_{o_3,o_4}\delta_{o_1,o_4} 
    + \left[\frac{U^2 \hat{\chi}_0(\bvec k + \bvec k^{\prime})}{1-U\hat{\chi}_0(\bvec k + \bvec k^{\prime})}\right]_{o_1,o_2}\delta_{o_1,o_4}\delta_{o_2,o_3} 
    + \left[\frac{U^3 \hat{\chi}_0(\bvec k - \bvec k^{\prime})^2}{1-U^2 \hat{\chi}_0(\bvec k - \bvec k^{\prime})^2}\right]_{o_1,o_2}\delta_{o_1,o_3}\delta_{o_2,o_4}.
\label{eq:Cvertex}
\end{split}
\end{align}
The effective spin fluctuation vertex Eq.~\eqref{eq:Cvertex} is proportional to $(1-U\chi_0)^{-1}$ reminiscent of the magnetic instability that is reached when the Stoner condition at $U \to U_c$ is met. For $U \geq U_c$ the theory hence breaks down as magnetic fluctuations diverge and the system orders magnetically. To avoid numerical instabilities when calculating superconducting gap functions using the effective pairing vertex $V^{\text{PP}}_C$, we perform the matrix inversions $1/(1-U\chi_0)$ and $1/(1-U^2(\chi_0)^2)$ in Eq.~\eqref{eq:Cvertex} in the eigenspace of $\chi_0$ and add a small imaginary broadening constant $i\,\eta_\mathrm{FLEX}=50\,\mathrm{meV}$ to all the eigenvalues in the denominator. Note that the energy scale of this broadening constant is to be compared with (and has to be very small in regard to) the \emph{vertex} energy scales, i.e. $4-5 \,\mathrm{eV}$ and \emph{not the valley-flat band} energy scales.

As explained in the main text, local interactions are flat in momentum space such that both inter- and intravalley coupling are present. In conjunction with single layer ferromagnetic fluctuations, the intervalley exchange enhances order parameters that are (i) spin-triplet and (ii) change sign under a valley flip. We find that local interactions exclusively promote valley-singlet, spin-triplet $f$-wave order for all fillings around the VHS in ABCB stacked graphene. We can understand the prevailance of $f$-wave superconductivity from a miscroscopic picture when analyzing the effective vertex structure Eq.~\eqref{eq:Cvertex} as shown in \Cref{fig:vertices}. As known from the one-band Hubbard model~\cite{scalapino1987} the spin fluctuation vertex remains purely positive in momentum space such that the symmetry of the superconducting order parameter must exhibit a sign change at the respective Fermi surface. We note that even though the effective vertex peaks around $\Gamma$, see \Cref{fig:vertices} (a,b and subsequent columns), the effective vertex has substantial amplitude for momentum transfers that connect the two valleys $K^{(\prime)}$. Therefore, the superconducting condensate can minimize its free energy by changing sign between the valleys leading to dominante $f$-wave SC order. At the same time, the sub-leading instability within the $x$RPA is of $p_{x,y}$-wave type, see \Cref{fig:fig4} (b) in the main text. The $p_{x,y}$ spin-triplet instability is an anti-symmetric basis function of the $E$ irreducible representation of $C_{3v}$ in momentum space. Similar to the $d$RPA results, the $p_{x,y}$-wave is driven by fluctuations that are centered at $\Gamma$. We hence note that the delicate interplay of short and long-ranged interactions, which we resolve within our orbital-resolved RPA, drive either order parameter.

\subsection{Long-Ranged Interactions}
Pairing mediated by screened Coulomb interactions can be accounted for by the effective pairing vertex
\begin{equation}
    \left [ V^{\text{PP}}_D \right ]_{o_1o_2o_3o_4}(\bvec k, \bvec k') = \left [ \hat{V}^{O}(\bvec k - \bvec k') \frac{1}{\mathbb{1} + 2  \hat{\chi}_{0}(\bvec k - \bvec k') \hat{V}^{O}(\bvec k - \bvec k')} \right ]_{o_1o_2} \delta_{o_1 o_3} \delta_{o_2 o_4},
\label{eq:Dvertex}
\end{equation}
where $\hat{V}^{O}(\bvec q) = V^{O}_{o_1o_2}(\bvec q)$ is the Fourier transform of the full Coulomb interaction and the factor of 2 in the denominator arises due to the internal trace over spin indices in the direct particle-hole channel. The latter is a direct consequence of the $SU(2)$ symmetry in the normal-state Hamiltonian. As described in the main text, we use a realistic (orbital-resolved) ``Ohno'' interaction profile~\cite{ohno1964some,wehling2011strength} 
\begin{equation}
    V^{O}(\bvec r) = \frac{Ua}{\sqrt{a^2+\bvec r^2}}e^{-|\bvec r|/d},
\label{eq:ohno_interaction-supp}
\end{equation}
with realistic parameters $a = 0.3 a_0$ and $d = 200 a_0$ (where $a_0 = 2.46\,\text{\AA}$ is the graphene lattice constant). The parameter $d$ controls the external screening from, e.g., metallic gates or the substrate, $a$ controls the atomic scale decay of the Coulomb interaction in real space, and $U$ sets the on-site (Hubbard) interaction strength. In particular, the `Ohno'' interaction profile ensures consistent treatment of long- and short-ranged terms of the Coulomb interaction. We further checked that our results do not change when varying the screening parameter in the realistic range of $d=50 \dots 200 a_0$, see Section~\ref{sec:parameters_SC}. 

As argued in the main text, order parameters with an intravalley sign change are generically favored by the screened Coulomb interaction for realistic interaction parameters as screening induced pairing Eq.~\eqref{eq:vertex_d} is most relevant for momenta $\bvec{q}_D$ where the polarization function $\hat{\chi}_0(\bvec{q}_D)$ peaks. To underline this argument, we show the momentum structure of the effective pairing vertex mediated by screened Coulomb interactions in \Cref{fig:vertices}. Indeed, we observe that opposite to the $x$RPA approach no amplitude exists at momentum transfers $\bvec q_D = K^{(\prime)}$ that connects the two valley.

\subsection{Combined mechanism for long- and short-ranged interactions}
\begin{figure}
    \centering
    \input{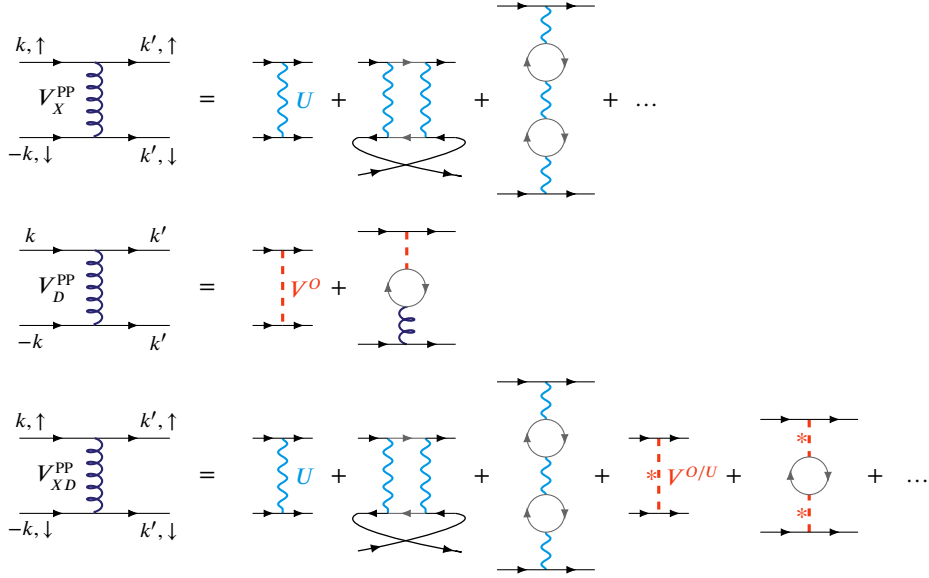}
    \caption{Diagrams contributing to the effective pairing vertex for different mechanims within the random phase approximations (RPA). The effective scattering vertex for Cooper pairs with momenta $(\bvec k, - \bvec k)$ and opposite spins for $x$RPA (upper row), $d$RPA (middle row) and $xd$RPA (lower row). Blue interaction lines indicate local Hubbard interactions, whereas the red interaction contains the full long-ranged (screened) Coulomb interaction with the Fourier transform $V^{O}(\bvec q)$ of the Ohno interaction profile Eq.~\eqref{eq:ohno_interaction}. The $x$RPA picks up contributions from the exchange as well as the direct particle-hole channel that contains an even number of loops between the in- and out-going legs. The $xd$RPA contains all diagrams from the $V^{\text{PP}}_X$ vertex, but additionally resums the long-ranged Coulomb interaction in the direct particle-hole channel. As we only consider scattering of Cooper pairs with opposite spin, the local interaction must be removed from the full Coulomb interaction $V^{O/U}$ in order to preserve the spin configuration of the effective vertex $V^{\text{PP}}_{XD}$. All diagramatic contributions are resummed to infinite order in the bare interactions, which is denoted by the dots or the self-consistent formulation of diagrams, respectively.}
    \label{fig:feynman-diagrams}
\end{figure}

Due to the different orbital- and momentum structures in the exchange and direct particle-hole channel, it is in general not possible to resum long-ranged interactions in the former channel. This is because the Fourier transform of the Ohno interaction $V^O_{o_1 o_2}(\bvec q_D)$ carries a channel-specific transfer momentum $\bvec q_D = \bvec k - \bvec k'$ which is native in the direct particle-hole channel (each interaction line in \Cref{fig:feynman-diagrams} carries transfer momentum $\bvec q$), but non-native in the exchange channel. Eq.~\eqref{eq:Cvertex} shows that the native momentum transfer in the exchange channel is $\bvec q_C = \bvec k + \bvec k'$ such that a simple RPA-like resummation with a single momentum transfer can no longer be achieved. In fact, it becomes necessary to unravel the full momentum and orbital dependence of the initial vertex and the susceptibilites, i.e. compute and store $\small V^O_{o_1 o_2 o_3 o_4}(\bvec k, \bvec k', \bvec q)$ and $\small \chi^0_{o_1 o_2 o_3 o_4}(\bvec k, \bvec q)$. To resolve the small Fermi surface patches in multilayer graphene systems, we sample the entire BZ with $N_{\bvec k} = 7200^2$ point and it hence becomes numerically unfeasible to obtain the effective pairing vertex. Therefore, we propose an alternative approach which only accounts for local interactions in the exchange channel and takes the full long-ranged interaction in the direct particle-hole channel, i.e. the best of the previous approaches that can be handled numerically.

As local interactions are accounted for in the exchange channel only, their contribution can be simply ported from the $x$RPA. For the direct particle-hole channel we would like to take the full long-ranged Coulomb interaction as in the $d$RPA. This can not be done trivially as we already know from the $x$RPA appraoch that the resummation within RPA of local interactions must be constrained such that only an even number of loops occur between the in- and out-going legs of the opposite spin vertex Eq.~\eqref{eq:Cvertex}. We can solve this problem by writing the spin-dependence of the susceptibility and the initial vertex explicitly and constrain the effective pairing vertex to the configuration $(\uparrow \downarrow \uparrow \downarrow)$ afterwards. To this end we define
\begin{align}
\begin{split}
    \tilde{\chi}^{s_1 s_2}_{o_1 o_2}(\bvec q) &= \chi^0_{o_1 o_2}(\bvec q) \delta_{s_1s_2} \\
    \tilde{V}^{s_1 s_2}_{o_1 o_2}(\bvec q) &= V^{O/U}_{o_1 o_2}(\bvec q) \delta_{s_1s_2} + U \delta_{s_1 \bar{s}_2},
\end{split}
\end{align}
where $V^{O/U}_{o_1 o_2}(\bvec q)$ is the Fourier transform of the Coulomb interaction, but without the local Hubbard-$U$ term. The renormalized interaction within RPA is hence given by 
\begin{equation}
\begin{split}
    \left [ \tilde{V}^{\text{PP}}_{XD} \right ]^{s_1s_2s_1s_2}_{o_1o_2o_3o_4}(\bvec k, \bvec k') &=  \left[ \hat{\tilde{V}}(\bvec k - \bvec k') \frac{1}{\mathbb{1} + \hat{\tilde{\chi}}(\bvec k - \bvec k') \hat{\tilde{V}}(\bvec k - \bvec k') } \right ]_{(s_1o_1)(s_2o_2)} \delta_{o_1 o_3} \delta_{o_2 o_4} \\
    \left [ V^{\text{PP}}_{XD} \right ]^{\uparrow \downarrow \uparrow \downarrow}_{o_1o_2o_3o_4}(\bvec k, \bvec k') &= \left [ \tilde{V}^{\text{PP}}_{XD} \right ]^{\uparrow \downarrow \uparrow \downarrow}_{o_1o_2o_3o_4}(\bvec k, \bvec k')
    +\left[\frac{U^2 \hat{\chi}_0(\bvec k + \bvec k^{\prime})}{1-U\hat{\chi}_0(\bvec k + \bvec k^{\prime})}\right]_{o_1,o_2}\delta_{o_1,o_4}\delta_{o_2,o_3} 
\end{split}
\label{eq:CDvertex}
\end{equation}
In the last step, we have restricted the vertex to the opposite spin configuration ($\uparrow \downarrow \uparrow \downarrow$) such that we recover all terms of the direct particle-hole channel that were present in the $x$RPA and the $d$RPA.

\subsection{Linearized gap equation}
To obtain information about the pairing symmetry of the underlying Cooper pairs as well as the superconducting coupling constant, i.e. the critical temperature of the superconducting transition, we solve a linearized gap equation in the projected subspace containing the bands in an energy window $W$ around the Fermi level
\begin{equation}
    \lambda_{\text{SC}}  \Delta_{b}(\bvec k) = - \frac{1}{N_{\bvec k}} \sum_{\bvec k' b'} \hat{V}^{\text{PP}}_{a, bb'}(\bvec k, \bvec k') \chi^{\text{PP}}_{b'}(\bvec k') \Delta_{b'}(\bvec k'), \quad a \in \{X,D,XD\}
\label{eq:lin_gap}
\end{equation}
where $\hat V^\mathrm{PP}_{bb'}$ and $\chi^\mathrm{PP}_{b'}$ denote the band-projected effective pairing vertex and particle-particle susceptibility, respectively. The largest eigenvalue $\lambda_\mathrm{SC}>0$ will lead to the highest transition temperature $T_c$ and the corresponding eigenfunction $\Delta(\bvec{k}, b)$ determines the symmetry of the gap, which can be classified according to the irreducible representations of the point group of the normal-state Hamiltonian. The band-projected quantities in Eq.~\eqref{eq:lin_gap} are defined as
\begin{align}
\label{eq:band_projections}
    \hat{V}^{\text{PP}}_{bb'}(\bvec k, \bvec k') &= \sum_{o_1 \dots o_4} u_{o_1,b}^*(\bvec k) u_{o_2,b}^*(-\bvec k) V^{\text{PP}}_{o_1,o_2,o_3,o_4}(\bvec k, \bvec k^{\prime}) u_{o_3,b}^{\phantom *}(\bvec k^{\prime}) u_{o_4,b}^{\phantom *}(-\bvec k^{\prime}) \,, \\
    \chi^{\text{PP}}_{b}(\bvec k) &=  \frac{n_F(\epsilon_b(\bvec k))-n_F(-\epsilon_b(-\bvec k))}{\epsilon_b(\bvec k) + \epsilon_b(-\bvec k)} \,. 
\end{align}
In ABCB, only a single band contributes to the Fermi surface on the hole-doped side in the vicinity of the VHS. Therefore, the linearized gap equation Eq.~\eqref{eq:lin_gap} can effectively be constrained to only capture points on the Fermi surface within the respective band, i.e. only intraband pairing of electrons becomes significant. To this end, we project the effective pairing vertex from orbital to band space Eq.~\eqref{eq:band_projections}. The vertex $\hat{V}^{\text{PP}}_{bb'}(\bvec k, \bvec k')$ hence describes scattering of Cooper pairs with momenta ($\bvec k, -\bvec k$) on Fermi surface patch $\mathcal{S}_b$ to ($\bvec k', -\bvec k'$) on Fermi surface patch $\mathcal{S}_{b'}$. Moreover, we can restrict the momentum dependence of the vertex to points on the Fermi surface. In fact, the problem of setting up the pairing vertex reduces to determine proper points on the Fermi surface, i.e. momentum points that satisfy $\mu_{\text{min}} < \epsilon_b(\bvec k) <\mu_{\text{max}}$. To treat this cutoff problem consistently, we choose $\mu_{\text{min}} = -5$ meV and $\mu_{\text{max}} = -2$ meV to capture all fillings on the hole-doped side for which we analyze the superconducting instabilities. In particular, we ensure that Fermi surface broadening may not be smaller than thermal broadening in the system: $\eta_{\text{FS}}>T$. The explicit contraction of the effective pairing vertex with the particle-particle loop Eq.~\eqref{eq:lin_gap} will suppress all contributions that are not in the immediate vicinity of the Fermi surface $\propto \epsilon^{-1}$.

\subsection{Temperature and Screening dependence of the superconducting order parameter}
\label{sec:parameters_SC}

In this section, we show how the superconducting instabilities discussed in the manuscript are subject to changes in temperature and the screening parameter $d$. 

\begin{figure*}
    \centering
    \includegraphics[width=0.5\textwidth]{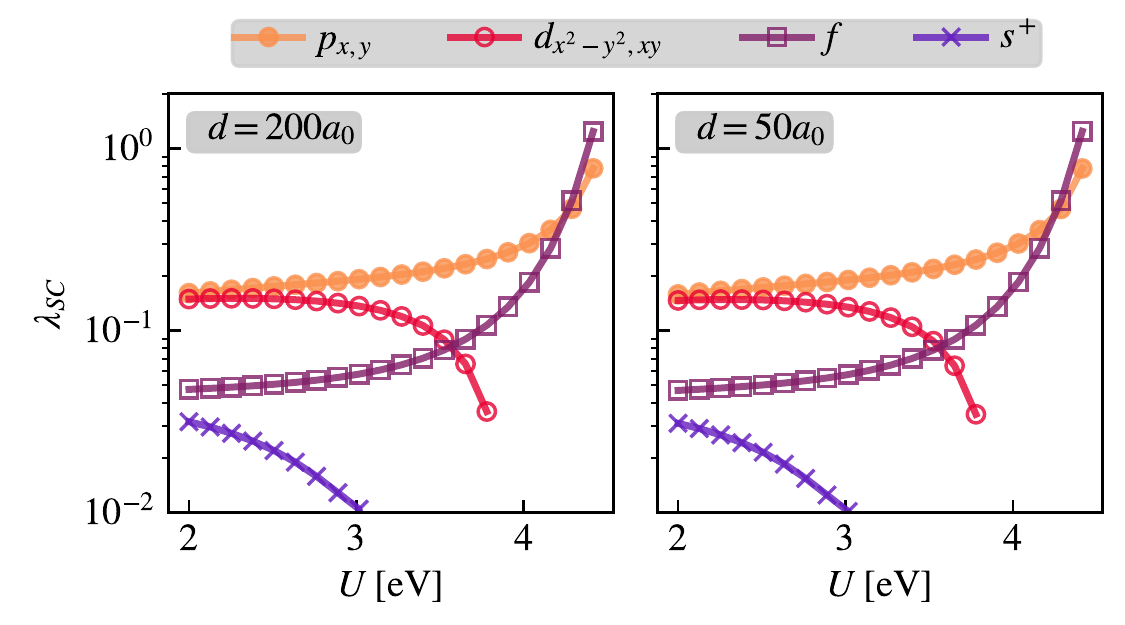}
    \caption{Screening dependence of the superconducting order within the $xd$RPA mechanism. Changing the screening parameter from $d=200 a_0$ to $d=50a_0$ leads to almost identical superconducting coupling constants $\lambda_{SC}$ and leaves the symmetry of the leading SC order paramter untouched.}
    \label{fig:sc_vs_d}
\end{figure*}

First, we study the robustness of our results when varying the screening parameter $d$. We note that screening of the long-ranged Coulomb interaction may in general also depend on environmental screening by e.g. metallic gates or the substrate. Our results are summarized in \Cref{fig:sc_vs_d}, where we show the superconducting coupling constant $\lambda_{SC}$ as function of the on-site Hubbard-$U$ at $\mu = \mu_{\text{VHS}}$ for different screening lengths of $d=50a_0, 200 a_0$ within the $xd$RPA approach. Apparently, all qualitative features remain robust when increasing the screening to $d = 50 a_0 \approx 10 $ nm the such that we conclude that screening of the long-ranged tail of the Coulomb interaction does not affect the formation of superconducting instabilities notably.

Next, we study the dependence of the superconducting coupling constant $\lambda_{SC}$ on the temperature broadening $T$ that enters the calculations of the particle-hole (particle-particle) susceptibilities. As electronic states are smeared in an energy window $\sim T$ around the Fermi surface due to the appearance of Fermi functions in Eq.~\eqref{eq:supp-loop} (Eq.~\eqref{eq:band_projections}), this also affects the formation of magnetic order and thus the values of the superconducting coupling constant $\lambda_{SC}$. Within the $x$RPA approach the critical Stoner value $U_c$ decreases with temperature as already demonstrated in \Cref{fig:chi-temperatures}. As superconductivity mediated by spin fluctuation is enhanced by the Stoner renormalization factor $(1-U \hat{\chi}_0)^{-1}$, we observe that the SC coupling constant $\lambda_{SC}$ is enhanced particularly at the VHS, where we approach the limit $U \to U_c$. At $U=4.5$ eV, we see that for $T = 0.25$ K, we already entered the Stoner phase as $\lambda_{SC}$ drops to zero, whereas at higher temperatures superconductivity prevails. Fro the $d$RPA we also observe a successive increase of $\lambda_{SC}$ by lowering the temperature. Especially the peak close to the valence band edge is increased significantly when lowering the temperature. The behaviour of the combined $xd$RPA approach can be deduced from the effects of the $x$RPA and $d$RPA separately.

\begin{figure*}
    \centering
    \includegraphics[width=0.75\textwidth]{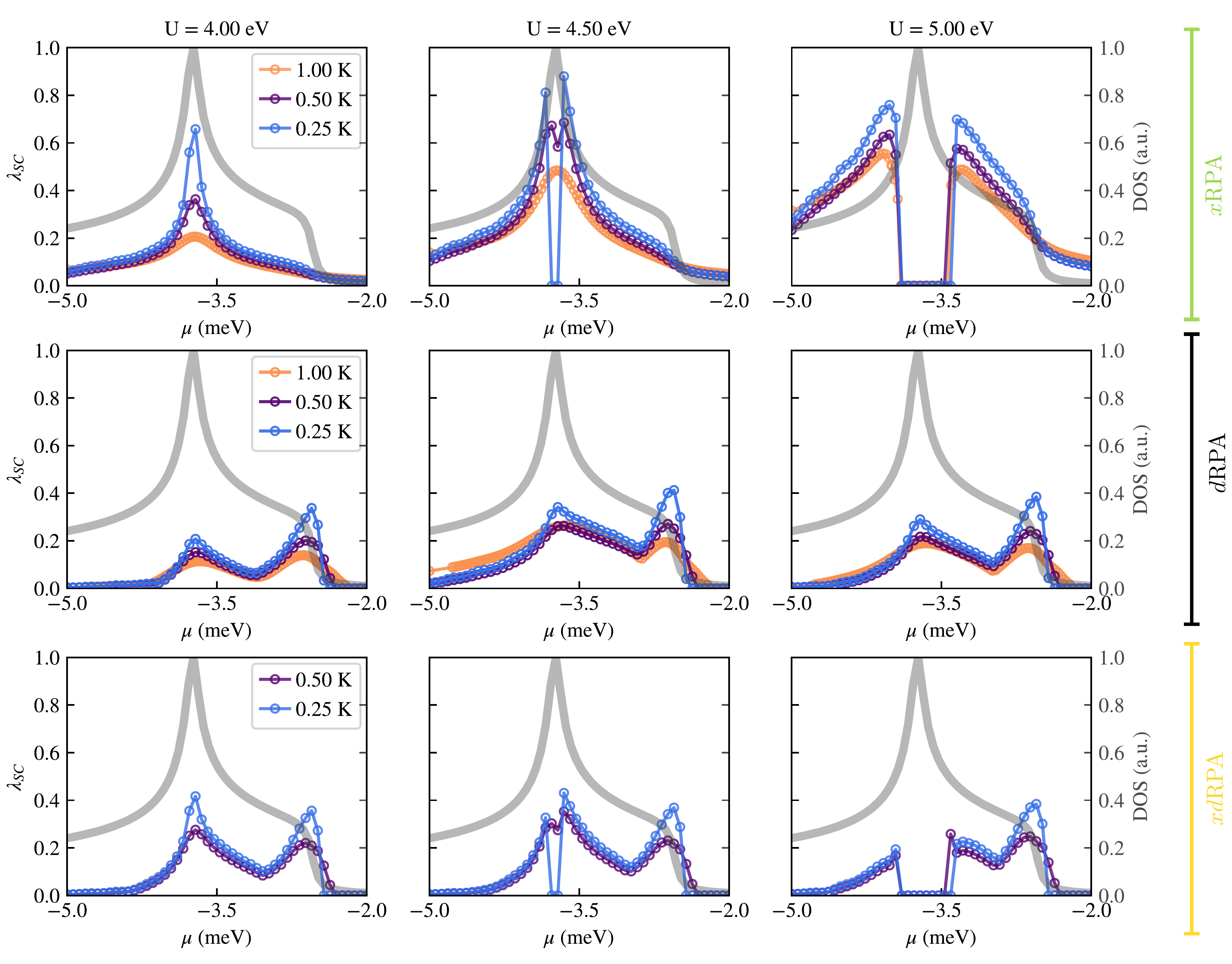}
    \caption{Temperature dependence for the different superconducting mechanisms $x$RPA, $d$RPA and $xd$RPA. Each panel shows the leading superconducting coupling constant $\lambda_{SC}$ as function of chemical potential $\mu$ on the hole-doped side of ABCB. Different columns display the results for different values of the local Hubbard-$U$ interactions, whereas different rows show the different mechanisms based on spin fluctuations ($x$RPA), screened Coulomb interactions ($d$RPA) and a combined mechanisms ($xd$RPA) that captures both long and short ranged Coulomb interactions.}
    \label{fig:sc_T}
\end{figure*}

\end{document}